\documentclass[aps,footinbib,prb,amsmath,amssymb,twocolumn,showpacs, superscriptaddress,10pt]{revtex4-1}
\usepackage[english]{babel}
\usepackage{braket}
\usepackage[table]{xcolor}
\usepackage{textcomp}
\usepackage{amsmath}
\usepackage{amssymb}
\usepackage{graphicx} 
\usepackage{}
\usepackage{pgf}
\usepackage{hyperref}
\usepackage{epstopdf}

\newcommand{\be}{\begin{eqnarray}}
\newcommand{\ee}{\end{eqnarray}}

\def\bra#1{\langle #1 |}
\def\ket#1{| #1 \rangle}

\definecolor{john}{rgb}{0.3,0.3,0.8}

\definecolor{alex}{rgb}{0,0.5,0.9}

\begin{document}

\title{Dynamical phase transition in the 1D-transverse field Ising chain characterized by the transverse magnetization spectral function}
\author{Giulia Piccitto} 
\affiliation{SISSA - International School for Advanced Studies, via Bonomea 265, 34136 Trieste, Italy}
\affiliation{Dipartimento di Fisica ed Astronomia, Universit\`a di Catania, Via Santa Sofia 64, 95123 Catania, Italy}
\author{Alessandro Silva}
\affiliation{SISSA - International School for Advanced Studies, via Bonomea 265, 34136 Trieste, Italy}

\begin{abstract}
	
	We study the response of a Quantum Ising Chain to transverse field oscillations in the asymptotic state attained after a quantum quench. We show that for quenches across a quantum phase transition the dissipative part of the response at low frequencies is negative, corresponding to energy emission up to a critical frequency $\omega^*$. The latter is found to be connected to the time period $t^*$ of the singularities  in the Loschmidt echo ($t^*=2\pi/'\omega^*$) signalling the presence of a dynamical phase transition.  This result suggests that linear response experiment can be used to detect this kind of phenomena.
\end{abstract}

\pacs{05.30.Rt, 64.60.Ht, 75.10.Jm}

\maketitle

\section{Introduction}

Recent improvements in experimental techniques \cite{art:Gring(2012), art:Smith(2013), art:Langen(2013), art:Langen(2015)} have given access to the quantum mechanical simulation of the dynamics of isolated, interacting quantum many body systems with a variety of platforms, such as cold atoms in optical lattices and ion traps \cite{art:Polkovnikov(2011), art:Gogolin(2011)}. 
An isolated system can be taken out of equilibrium in many ways: the commonest protocol is known as quantum quench, i.e. an instantaneous change of one of the parameters of the Hamiltonian starting from a pre-quench ground state. The stationary state attained after a quench, in particular, gives a great deal of information on the system, on its integrability \cite{art:Langen(2015), art:Gambassi(2012), art:Calabrese(2011), art:Calabrese(2012), art:Calabrese(2012):2, art:Hessler(2016), art:Hess(2017)} or ergodicity \cite{art:Rigol(2008), art:Srednicki(1994)}, as well as on the nature of its quasiparticles. 

Stationary states may display critical behavior as a function of the quench parameters \cite{art:Zhang(2017)}. When this is evinced from an order parameter we encounter a type of criticality naturally reducible to the Landau paradigm~\cite{art:Gambassi_C(2011),art:Sciolla(2010)} in classical systems. 
A second type of transition\cite{art:Heyl(2013)}, more subtle and ephemeral, can be observed by studying the characteristics of the time evolved state itself $\ket{ \psi (t)} = e^{-i H t} \ket{\psi_0}$. When starting from a broken symmetry ground state the probability to remain in the ground state manifold displays time-periodic cusp singularities \cite{art:Heyl(2014), art:Jurcevic(2017)} for a range of quench parameters which turns out to be connected with the emergence of stationary state criticality \cite{art:Knap(2018)}.
Singularities may emerge also in the return amplitude  $ \braket{\psi(t)|\psi_0}$ (Loschmidt echo) when the time evolved state is orthogonal to the initial one. Despite the observation of such singularities in quenches across an Ising quantum critical point \cite{art:Knap(2018)} and in the collapse and revival dynamics of superfluids \cite{art:Lacki(2018), art:Fogarty(2017)}, their significance and the conditions of their emergence have been under investigation since their conception \cite{art:Heyl(2018), art:Halimeh(2017), art:Halimeh(2017)2, art:Halimeh(2018)}.

The purpose of this work is to suggest a connection between the DPT and the emergence of a peculiar response of a system to perturbations which could be used to detect them. Focusing on the dynamics of the Transverse Field Ising Model (TFIM), where DPT are expected for quenches across the quantum critical point, we will show that whenever a DPT occurs the system displays an anomalous low frequency response to transverse field oscillations in the steady state. Unlike perturbed system in equilibrium, the system emits energy at a finite rate and the dissipative part of the response function is negative\cite{art:Rossini(2014)}. Interestingly this anomalous response occurs for perturbations up to a critical frequency $\omega^*$ which is in direct correspondence with the time period $t^*=2\pi/\omega^*$ of 
the singularities of the return probability. 

The outline of the rest of the paper is as follows: in Section ~\ref{sec:2} the TFIM and the quench protocol are introduced. After a brief summary of the concept of DPTs and linear response theory in the context of quenches, the calculation of the transverse magnetization spectral function is presented. In Section ~\ref{sec:3} the results are presented and analyzed for different quenches. Finally we present our conclusions in Section ~\ref{sec:4}.

\section{The Model, Linear Response Theory and Dynamical Phase Transitions.}\label{sec:2}

The Hamiltonian we study is that of the TFIM given by
\begin{equation}
	\hat{H}(h) = - J \sum_i \hat{\sigma}_i^x \hat{\sigma}_{i+1}^x - h \sum_i \sigma^z_i,
\end{equation}
where $\hat{\sigma}^x_i$ and $\hat{\sigma}^z_i$ are Pauli matrices representing the $x$ and $z$ components of the spin at the $i$-th site. We will consider the ferromagnetic case only and set $J = 1$ and $\hbar = 1$. At equilibrium, this system exhibits a quantum phase transition at $h = 1$. It is well known that \cite{art:Barouch(1970), art:Barouch(1971), book:Sachdev(2000)} this model can be diagonalized by mapping the Hamiltonian onto a system of spinless fermions, and performing in the $k-$space a Bogoliubov rotation to end up with a system of free fermions $\gamma_k$ with dispersion $\varepsilon_k = 2 \sqrt{(h - \cos k)^2 + \sin^2 k}$ characterized by a gap $\Delta=2~| h-1|$.

Before discussing the characterization of the dynamics of this system in terms of a response function, let us briefly discuss in mathematical terms the concept of DPTs defined by Heyl {\it et al}~~\cite{art:Heyl(2013)}. DPTs primarily centred around the Loschmidt amplitude
\begin{equation}\label{amplitude}
\mathcal{G}(t)=\bra\Psi e^{-iHt}\ket\Psi
\end{equation}
that has been extensively studied under a number of guises in the past decades. 
This amplitude, following a Wick rotation i.e $z=it$, can be thought of as a boundary partition function ie $\mathcal{Z}(z)=\bra\Psi e^{-z H}\ket\Psi$  when $z\in\mathbb{R}$. Exploiting this mapping Heyl {\it et al} noticed that as the free energy density can be defined as $f(z)=-\lim_{L\rightarrow\infty}\frac{1}{L}\ln Z(z)$ then in direct analogy with standard statistical mechanics, the Fisher zeros in this boundary partition function (corresponding to singularities in $f(t)$) coalesce into lines which can cross the real axis. The repercussion of this behaviour in the temporal domain leads to the emergence of critical times $t^{*}_n$ where the so called rate function 
\begin{equation}\label{rate}
l(t)=-\frac{1}{L}\ln |\mathcal{G}(t)|^{2}
\end{equation}
displays non-analyticities.  
According to this definition of DPTs put forward in Ref.\onlinecite{art:Heyl(2013)}, the singularities identify points at which the time evolved state is orthogonal to the initial one.

In the case of a quench of the transverse field $h_i \rightarrow h$ in the TFIM across the quantum phase transition the singularities in the rate function $l(t)$ were found to occur at $t^*_n=t^*(n+1/2)$ and $t^*=\pi/\varepsilon_{k^*}$ where the momentum $k^*$ is determined by the condition $\cos(k^*)=(1+h_ih)/(h_i+h)$. The momentum $k^*$ marks the transition in the occupation of fermionic quasiparticle from {\it thermal} ($n_k<1/2$ for $|k|>k^*$) to {\it non-thermal} ($n_k>1/2$ for $|k|<k^*$). A population inversion in non-equilibrium systems opens up the possibility of extacting work in subsequent cyclical transformations~\cite{art:Allahverdyan(2004)}. In other words, while equilibrium systems heat up as a result of a periodic drive a population inversion opens up the possibility of energy emission, as observed previously in Ref.~(\onlinecite{art:Rossini(2014)}).   

A natural way to investigate these issues is to study the dissipative of linear response functions which, also in the case of quantum quenches, is naturally connected to the dynamic of energy absorbtion~\cite{art:Russomanno(2013)}.  In the present situation the simplest possible experiment could consist in studying  the linear response after a quench of the transverse field. We start with the Hamiltonian $\hat{H}(h_i)$ and the system prepared in its ground state $\ket{\psi_0}$. At $t = 0$ we change suddenly the value of the transverse field and we add a small periodic perturbation to break time and space translational invariance
\begin{equation}
	\hat{H}(h_i) \mapsto \hat{H}(h) + \hat{v}(t),
\end{equation}
with $\hat{v}(t) = \delta h \sum_j \sin (\omega_0 t) \sin (k_0 j) \sigma_j^z$ and $\delta h \ll 1$. 
Even in the nonequilibrium case~\cite{art:Rossini(2014)} the response function of the transverse magnetization can be written as
\begin{equation}
	\chi_{z z}(l, l', t, t') = -i \theta(t-t') \bra{\psi_0}[\hat{\sigma^z_l}(t), \hat{\sigma^z_{l'}}(t')]\ket{\psi_0}.
\end{equation}
Since the system is transnationally invariant then $\chi$ depends on $r=l-l'$ only. The out of equilibrium condition makes however the dependence on $t$ and $t'$ essential.
The latter can be described also in terms of Wigner coordinates 
\begin{equation}
	\begin{array}{cc}
		\tau = t-t',\\ \\
		T = \frac{t+t'}{2}.
	\end{array}
\end{equation}
Taking a Fourier transform with respect to the relative coordinates $\tau,r$ 
\begin{equation}
	\chi(q, \omega, T) = \int dr e^{i q r} \int d(\tau) e^{i \omega \tau} \chi(r, T, \tau),
\end{equation}
the response in the asymptotic stationary state can be obtained by performing an average with respect to the time $T$ 
\begin{equation}\label{spec}
	\overline{\chi(q, \omega)} = \lim_{T \to \infty} \frac{1}{T} \int_0^T \chi(q, \omega, T).
\end{equation}

In order to study the dynamics of energy asborbsion we have to compute the imaginary part of the time-averaged response $\bar{\chi}''$. The calculation of the latter
employs standard techniques: both the initial and the final Hamiltonian can be expressed in fermionic form following a Jordan Wigner transformation. Expressing the 
final quasiparticles $\gamma_f(k)$ in terms of the initial ones $\gamma_i(k)$ via a Bogoliubov rotation, the correlation functions of quadratic operators such as the transverse field
can be easily calculated. We obtain the spectral function
\begin{equation}\label{chi_equation}
\begin{aligned}
	\overline{\chi}(q, \omega)'' = - &\frac{2\pi}{L} \sum_k \delta( \varepsilon_{k - \frac{q}{2}} + \varepsilon_{k + \frac{q}{2}} + \omega) A\left(k, q\right) \\- &\frac{2\pi}{L} \sum_k \delta( \varepsilon_{k - \frac{q}{2}} + \varepsilon_{k + \frac{q}{2}} - \omega) B\left(k, q\right)\\ - &\frac{2\pi}{L} \sum_k \delta( \varepsilon_{k - \frac{q}{2}} - \varepsilon_{k + \frac{q}{2}} - \omega) C\left(k, q\right)\\ - &\frac{2\pi}{L} \sum_k \delta( \varepsilon_{k - \frac{q}{2}} - \varepsilon_{k + \frac{q}{2}} + \omega) D\left(k, q\right),
\end{aligned}
\end{equation}
where
\begin{equation}
	\begin{aligned}
		A\left(k, q\right) &= \sin^2 \theta_{k-\frac{q}{2}} \cos^2 \theta_{k +\frac{q}{2}} \left( \cos^2 \delta_{k + \frac{q}{2}} - \sin^2 \delta_{k-\frac{q}{2}}\right),\\
		B\left(k, q\right) &= \cos^2 \theta_{k-\frac{q}{2}} \sin^2 \theta_{k + \frac{q}{2}} \left( \cos^2 \delta_{k-\frac{q}{2}} - \sin^2 \delta_{k+\frac{q}{2}}\right),\\
		C\left(k, q\right) &= \sin^2 \theta_{k-\frac{q}{2}} \sin^2 \theta_{k + \frac{q}{2}}\left( \cos^2 \delta_{k + \frac{q}{2}} - \cos^2 \delta_{k-\frac{q}{2}}\right),\\
		D\left(k, q\right) &= \cos^2 \theta_{k-\frac{q}{2}} \cos^2 \theta_{k + \frac{q}{2}} \left( \cos^2 \delta_{k-\frac{q}{2}} - \cos^2 \delta_{k+\frac{q}{2}}\right).  
	\end{aligned}
\end{equation}
In the previous expression we have set $\delta_k = \theta_k^f - \theta_k^i$, being $\theta_k^f$ and $\theta_k^i$ the angles of the Bogoliubov rotation that diagonalize respectively the pre-quench and the post-quench Hamiltonian ($ 2 \theta_k = \frac{\sin k}{h - \cos k}$). 
We are now ready to explore more in detail the connection of energy dynamics resulting from a perturbation to DTPs.  

\begin{figure*}[t!]
	\centering
	\includegraphics[width = \textwidth]{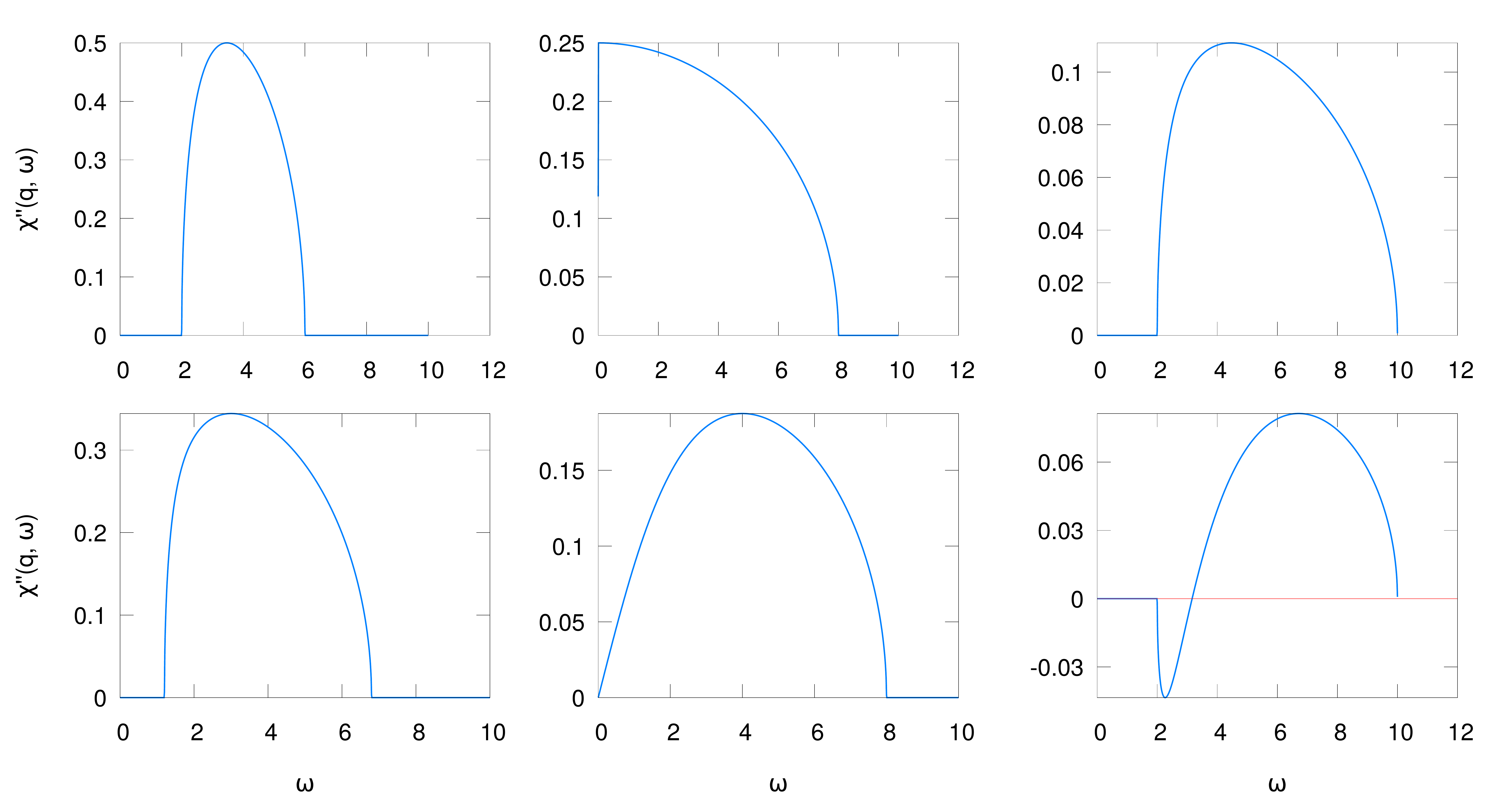}
	\caption{Top panels: transverse magnetization spectral function for an equilibrium Ising model in the presence of a transverse uniform perturbation for three different values of the transverse field $h = 0.5$ (left panel), $h = 1$ (central panel) and $h = 1.5$ (right panel). The spectrum of quasi particles is gapless only for $h = 1$.
	Bottom panels:Transverse magnetization spectral function for a nonequilibrium Ising model in presence of a transverse uniform perturbation for three different quenches: inside the ferromagnetic phase (left panel) from $h_i = 0.5$ to $h = 0.7$, toward the critical point from $h_i = 0.5$ to $h = 1$ (central panel) and across the transition point from $h_i = 0.5$ to $h = 1.5$ (right panel).}
	\label{chi}
\end{figure*}

\section{Results}\label{sec:3}

\begin{figure*}[t!]
	\centering
	\includegraphics[width = \textwidth]{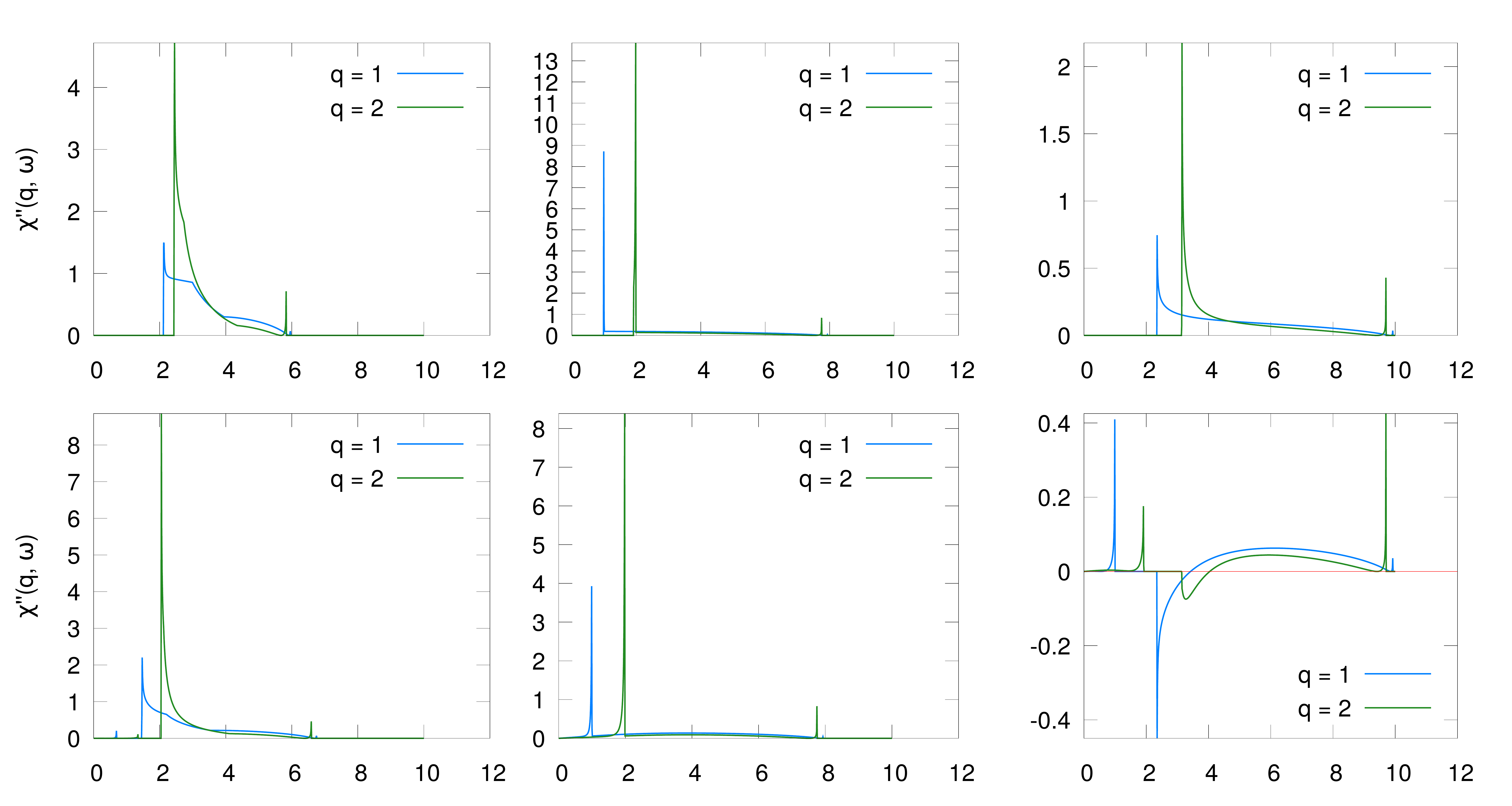}
	\caption{(color online) Top panels: transverse magnetization spectral function for an equilibrium Ising model in the presence of a transverse nonuniform perturbation for three different values of the transverse field $h = 0.5$ (left panel), $h = 1$ (central panel) and $h = 1.5$ (right panel). The spectrum of quasi particles is gapless only for $h = 1$. The two different colors correspond to two different values of the transferred momentum, blue line $q = 1$, green line $q = 2$.
	Bottom panels: transverse magnetization spectral function for a non equilibrium Ising model in presence of a transverse non uniform perturbation for three different quenches: inside the ferromagnetic phase (left panel) from $h_i = 0.5$ to $h = 0.7$, toward the critical point from $h_i = 0.5$ to $h = 1$ (central panel) and across the transition point from $h_i = 0.5$ to $h = 1.5$ (right panel). The two different colors correspond to two different values of the transferred momentum, blue line $q = 1$, green line $q = 2$.}
	\label{chi_2}
\end{figure*}

Let us first analyze the results we found in the case of uniform perturbation ($q = 0$), for positive frequencies, both for the equilibrium and for the nonequilibrium case.
Manipulating Eq. \eqref{chi_equation} it is easy to find an analytical expression for the transverse magnetization spectral function 
\begin{equation}
	\overline{\chi''(0, \omega)} = \chi''_{\text{eq}}(0, \omega)\  \frac{4\left(hh_i -(h+h_i)f(h, \omega)+1\right)}{\omega\sqrt{h_i^2-2h_if(h, \omega)+1}},
\end{equation}
where
\begin{equation}
	\chi''_{\text{eq}}(0, \omega) = \frac{\sqrt{1 - f(h, \omega)^2}}{h\omega},
\end{equation}
is the equilibrium one, and
\begin{equation}
	f(h, \omega) = \frac{1}{2h} \left(h^2 - \frac{\omega^2}{16} + 1 \right).
\end{equation}

The top panels of Fig. (\ref{chi}) show the equilibrium spectral function for the three different cases: ferromagnetic phase (upper left panel), critical point (upper right panel), paramagnetic phase (lower panel). Clearly the spectral function has support only at energies above the gap for two particle excitations $2 \Delta = 4|1-h|$, since only pairs can be emitted as a result of parity conservation. The same is true also in the case of a quench, shown in the bottom panels show for the three different quenches: inside the ferromagnetic phase (left panel) from $h_i = 0.5$ to $h = 0.7$, toward the critical point from $h_i = 0.5$ to $h = 1$ (central panel) and across the critical point from $h_i = 0.5$ to $h = 1.5$ (right panel).

 We notice that for quenches across the quantum phase transition the response function has negative imaginary part right above the gap $2\Delta$. This indicates that the system emits energy as a result of the coupling to the perturbation (\onlinecite{art:Russomanno(2013)}). As anticipated above, this is possible as a result of a population inversion: for fermions right above the gap the density of quasi-particles
\begin{equation}
		\braket{n_h(k)}_{h_i}= \frac{1- \cos 2\delta_k}{2}
	\end{equation}
is non thermal up to momenta satisfying the condition
\begin{equation}
	\cos 2 \delta_k = \frac{hh_i - (h + h_i) \cos k + 1}{\sqrt{h^2 -2h\cos k + 1}\sqrt{h_i^2 -2 h_i \cos k + 1}} \ge 0.
\end{equation}
It is easy to see that such region of population inversion exists if and only if only if $h < 1$ and $h_i > 1$ or viceversa. The resulting $k^* = \arccos \left(\frac{hh_i +1}{hh_i}\right)$  corresponds to an energy scale $\omega^* = 2 \varepsilon_k^*$ at which the spectral function vanishes too

\begin{equation}
\begin{aligned}
	&hh_i - \frac{(h + h_i)}{2h} \left( h^2 - \frac{\omega^2}{16} +1 \right) + 1 = 0 \\ &\iff \omega^2 = 16 \left( h^2 - 2h \left( \frac{h h_i + 1}{h + h_i}  \right) + 1 \right) = (2 \varepsilon_k)^2.
\end{aligned}
\end{equation}
As anticipated above $t^* = \frac{2\pi}{\omega^*}$ corresponds as expected to the period of the singularities in the Loschmidt echo signaling a dynamical phase transition according to Ref.(\onlinecite{art:Heyl(2013)}).

Let us now briefly consider how the scenario presented for $q=0$ is modified for finite $q$.  The results we found are shown in Fig. (\ref{chi_2}). In the top panels we show the transverse magnetization spectral function for an equilibrium Ising model in the presence of a transverse non uniform perturbation for three different values of the transverse field $h = 0.5$ (left panel), $h = 1$ (central panel) and $h = 1.5$ (right panel). The two different colors correspond to two different values of the transferred momentum, blue line $q = 1$, green line $q = 2$.
In the bottom panels we show the transverse magnetization spectral function for a non equilibrium Ising model in presence of a transverse non uniform perturbation for three different quenches: inside the ferromagnetic phase (left panel) from $h_i = 0.5$ to $h = 0.7$, toward the critical point from $h_i = 0.5$ to $h = 1$ (central panel) and across the transition point from $h_i = 0.5$ to $h = 1.5$ (right panel). The two different colors correspond to two different values of the transferred momentum, blue line $q = 1$, green line $q = 2$.
 What emerges is that in the presence of a non uniform perturbation the zeros of the spectral function move towards different frequencies. 
The same analysis done for the $q = 0$ case can be carried but the quasi particles density of population is now given by 
\begin{equation}
	\braket{n_h(k, q)}_{h_i} = \frac{1}{2} \left(\braket{n_h(k + \frac{q}{2})}_{h_i}+\braket{n_h(k-\frac{q}{2})}_{h_i}\right).
\end{equation}
The new condition for the population inversion is
\begin{equation}
\cos 2 \delta_{k + \frac{q}{2}} + \cos 2 \delta_{k- \frac{q}{2}} < 0
\end{equation} 
and it should be solved numerically to find the quench parameter for which some population inversion can be observed. This condition can also be used to find the value $q^*$ for which population inversion disappears. As an example, in case of a quench from $h_i = 0.5$ to $h = 1.5$ we find $q^* \sim 2.63$. A direct plot of the response function confirms that indeed for $q > q^*$ the spectral function is always positive in this case. 
Unlike the case $q=0$ in this case we have that the population inversion is ultimately not directly connected to the DPT, though ultimately it shares the same physical origin.

\section{Conclusions}\label{sec:4}
In this paper, we examined the spectral function of the transverse magnetization for a non equilibrium transverse field Ising model after a quantum quench. We have  shown that the population inversion that is behind the emergence of dynamical criticality can be directly detected in the dynamics of energy absorption from a periodic perturbation. It would be interesting to extend our analysis to other more complex models and make the connection between negative response functions and dynamical criticality, if possible, more general. We leave this to future endeavours.

\bibliography{biblio_articolo}

\begin{thebibliography}{33}%
\makeatletter
\providecommand \@ifxundefined [1]{%
 \@ifx{#1\undefined}
}%
\providecommand \@ifnum [1]{%
 \ifnum #1\expandafter \@firstoftwo
 \else \expandafter \@secondoftwo
 \fi
}%
\providecommand \@ifx [1]{%
 \ifx #1\expandafter \@firstoftwo
 \else \expandafter \@secondoftwo
 \fi
}%
\providecommand \natexlab [1]{#1}%
\providecommand \enquote  [1]{``#1''}%
\providecommand \bibnamefont  [1]{#1}%
\providecommand \bibfnamefont [1]{#1}%
\providecommand \citenamefont [1]{#1}%
\providecommand \href@noop [0]{\@secondoftwo}%
\providecommand \href [0]{\begingroup \@sanitize@url \@href}%
\providecommand \@href[1]{\@@startlink{#1}\@@href}%
\providecommand \@@href[1]{\endgroup#1\@@endlink}%
\providecommand \@sanitize@url [0]{\catcode `\\12\catcode `\$12\catcode
  `\&12\catcode `\#12\catcode `\^12\catcode `\_12\catcode `\%12\relax}%
\providecommand \@@startlink[1]{}%
\providecommand \@@endlink[0]{}%
\providecommand \url  [0]{\begingroup\@sanitize@url \@url }%
\providecommand \@url [1]{\endgroup\@href {#1}{\urlprefix }}%
\providecommand \urlprefix  [0]{URL }%
\providecommand \Eprint [0]{\href }%
\providecommand \doibase [0]{http://dx.doi.org/}%
\providecommand \selectlanguage [0]{\@gobble}%
\providecommand \bibinfo  [0]{\@secondoftwo}%
\providecommand \bibfield  [0]{\@secondoftwo}%
\providecommand \translation [1]{[#1]}%
\providecommand \BibitemOpen [0]{}%
\providecommand \bibitemStop [0]{}%
\providecommand \bibitemNoStop [0]{.\EOS\space}%
\providecommand \EOS [0]{\spacefactor3000\relax}%
\providecommand \BibitemShut  [1]{\csname bibitem#1\endcsname}%
\let\auto@bib@innerbib\@empty
\bibitem [{\citenamefont {Gring}\ \emph {et~al.}(2012)\citenamefont {Gring},
  \citenamefont {Kuhnerta}, \citenamefont {Langen}, \citenamefont {Kitagawa},
  \citenamefont {Rauer}, \citenamefont {Schreitl}, \citenamefont {Mazets},
  \citenamefont {Smith}, \citenamefont {Demler},\ and\ \citenamefont
  {Schmiedmayer}}]{art:Gring(2012)}%
  \BibitemOpen
  \bibfield  {author} {\bibinfo {author} {\bibfnamefont {M.}~\bibnamefont
  {Gring}}, \bibinfo {author} {\bibfnamefont {M.}~\bibnamefont {Kuhnerta}},
  \bibinfo {author} {\bibfnamefont {T.}~\bibnamefont {Langen}}, \bibinfo
  {author} {\bibfnamefont {T.}~\bibnamefont {Kitagawa}}, \bibinfo {author}
  {\bibfnamefont {B.}~\bibnamefont {Rauer}}, \bibinfo {author} {\bibfnamefont
  {M.}~\bibnamefont {Schreitl}}, \bibinfo {author} {\bibfnamefont
  {L.}~\bibnamefont {Mazets}}, \bibinfo {author} {\bibfnamefont {D.~A.}\
  \bibnamefont {Smith}}, \bibinfo {author} {\bibfnamefont {E.}~\bibnamefont
  {Demler}}, \ and\ \bibinfo {author} {\bibfnamefont {J.}~\bibnamefont
  {Schmiedmayer}},\ }\href@noop {} {\bibfield  {journal} {\bibinfo  {journal}
  {Science}\ }\textbf {\bibinfo {volume} {337}},\ \bibinfo {pages} {1318}
  (\bibinfo {year} {2012})}\BibitemShut {NoStop}%
\bibitem [{\citenamefont {Smith}\ \emph {et~al.}(2013)\citenamefont {Smith},
  \citenamefont {Gring}, \citenamefont {Langen}, \citenamefont {Kuhnert},
  \citenamefont {Rauer}, \citenamefont {Geiger}, \citenamefont {Kitagawa},
  \citenamefont {Mazets}, \citenamefont {Demler},\ and\ \citenamefont
  {Schmiedmayer}}]{art:Smith(2013)}%
  \BibitemOpen
  \bibfield  {author} {\bibinfo {author} {\bibfnamefont {D.~A.}\ \bibnamefont
  {Smith}}, \bibinfo {author} {\bibfnamefont {M.}~\bibnamefont {Gring}},
  \bibinfo {author} {\bibfnamefont {T.}~\bibnamefont {Langen}}, \bibinfo
  {author} {\bibfnamefont {M.}~\bibnamefont {Kuhnert}}, \bibinfo {author}
  {\bibfnamefont {B.}~\bibnamefont {Rauer}}, \bibinfo {author} {\bibfnamefont
  {R.}~\bibnamefont {Geiger}}, \bibinfo {author} {\bibfnamefont
  {T.}~\bibnamefont {Kitagawa}}, \bibinfo {author} {\bibfnamefont
  {I.}~\bibnamefont {Mazets}}, \bibinfo {author} {\bibfnamefont
  {E.}~\bibnamefont {Demler}}, \ and\ \bibinfo {author} {\bibfnamefont
  {J.}~\bibnamefont {Schmiedmayer}},\ }\href@noop {} {\bibfield  {journal}
  {\bibinfo  {journal} {New Journal of Physics}\ }\textbf {\bibinfo {volume}
  {15}},\ \bibinfo {pages} {075011} (\bibinfo {year} {2013})}\BibitemShut
  {NoStop}%
\bibitem [{\citenamefont {Langen}\ \emph {et~al.}(2013)\citenamefont {Langen},
  \citenamefont {Geiger}, \citenamefont {Kuhnert}, \citenamefont {Rauer},\ and\
  \citenamefont {Schmiedmayera}}]{art:Langen(2013)}%
  \BibitemOpen
  \bibfield  {author} {\bibinfo {author} {\bibfnamefont {T.}~\bibnamefont
  {Langen}}, \bibinfo {author} {\bibfnamefont {R.}~\bibnamefont {Geiger}},
  \bibinfo {author} {\bibfnamefont {M.}~\bibnamefont {Kuhnert}}, \bibinfo
  {author} {\bibfnamefont {B.}~\bibnamefont {Rauer}}, \ and\ \bibinfo {author}
  {\bibfnamefont {J.}~\bibnamefont {Schmiedmayera}},\ }\href@noop {} {\bibfield
   {journal} {\bibinfo  {journal} {Nature Physics}\ }\textbf {\bibinfo {volume}
  {9}},\ \bibinfo {pages} {640} (\bibinfo {year} {2013})}\BibitemShut {NoStop}%
\bibitem [{\citenamefont {Langen}\ \emph {et~al.}(2015)\citenamefont {Langen},
  \citenamefont {Erne}, \citenamefont {Geiger}, \citenamefont {Rauer},
  \citenamefont {Schweigler}, \citenamefont {Kuhnert}, \citenamefont
  {Rohringer}, \citenamefont {Mazets}, \citenamefont {Gasenzer},\ and\
  \citenamefont {Schmiedmayer}}]{art:Langen(2015)}%
  \BibitemOpen
  \bibfield  {author} {\bibinfo {author} {\bibfnamefont {T.}~\bibnamefont
  {Langen}}, \bibinfo {author} {\bibfnamefont {S.}~\bibnamefont {Erne}},
  \bibinfo {author} {\bibfnamefont {R.}~\bibnamefont {Geiger}}, \bibinfo
  {author} {\bibfnamefont {B.}~\bibnamefont {Rauer}}, \bibinfo {author}
  {\bibfnamefont {T.}~\bibnamefont {Schweigler}}, \bibinfo {author}
  {\bibfnamefont {M.}~\bibnamefont {Kuhnert}}, \bibinfo {author} {\bibfnamefont
  {W.}~\bibnamefont {Rohringer}}, \bibinfo {author} {\bibfnamefont
  {I.}~\bibnamefont {Mazets}}, \bibinfo {author} {\bibfnamefont
  {T.}~\bibnamefont {Gasenzer}}, \ and\ \bibinfo {author} {\bibfnamefont
  {J.}~\bibnamefont {Schmiedmayer}},\ }\href@noop {} {\bibfield  {journal}
  {\bibinfo  {journal} {Science}\ }\textbf {\bibinfo {volume} {348}},\ \bibinfo
  {pages} {207} (\bibinfo {year} {2015})}\BibitemShut {NoStop}%
\bibitem [{\citenamefont {Polkovnikov}\ \emph {et~al.}(2011)\citenamefont
  {Polkovnikov}, \citenamefont {Sengupta}, \citenamefont {Silva},\ and\
  \citenamefont {Vengalattore}}]{art:Polkovnikov(2011)}%
  \BibitemOpen
  \bibfield  {author} {\bibinfo {author} {\bibfnamefont {A.}~\bibnamefont
  {Polkovnikov}}, \bibinfo {author} {\bibfnamefont {K.}~\bibnamefont
  {Sengupta}}, \bibinfo {author} {\bibfnamefont {A.}~\bibnamefont {Silva}}, \
  and\ \bibinfo {author} {\bibfnamefont {M.}~\bibnamefont {Vengalattore}},\
  }\href@noop {} {\bibfield  {journal} {\bibinfo  {journal} {Rev. Mod. Phys.}\
  }\textbf {\bibinfo {volume} {83}},\ \bibinfo {pages} {863} (\bibinfo {year}
  {2011})}\BibitemShut {NoStop}%
\bibitem [{\citenamefont {Gogolin}\ and\ \citenamefont
  {Eisert}(2016)}]{art:Gogolin(2011)}%
  \BibitemOpen
  \bibfield  {author} {\bibinfo {author} {\bibfnamefont {C.}~\bibnamefont
  {Gogolin}}\ and\ \bibinfo {author} {\bibfnamefont {J.}~\bibnamefont
  {Eisert}},\ }\href@noop {} {\bibfield  {journal} {\bibinfo  {journal} {Rep.
  Prog. Phys.}\ }\textbf {\bibinfo {volume} {79}},\ \bibinfo {pages} {056001}
  (\bibinfo {year} {2016})}\BibitemShut {NoStop}%
\bibitem [{\citenamefont {Foini}\ \emph {et~al.}(2012)\citenamefont {Foini},
  \citenamefont {Cugliandolo},\ and\ \citenamefont
  {Gambassi}}]{art:Gambassi(2012)}%
  \BibitemOpen
  \bibfield  {author} {\bibinfo {author} {\bibfnamefont {L.}~\bibnamefont
  {Foini}}, \bibinfo {author} {\bibfnamefont {L.~F.}\ \bibnamefont
  {Cugliandolo}}, \ and\ \bibinfo {author} {\bibfnamefont {A.}~\bibnamefont
  {Gambassi}},\ }\href@noop {} {\bibfield  {journal} {\bibinfo  {journal} {J.
  Stat. Mech.}\ }\textbf {\bibinfo {volume} {2012}},\ \bibinfo {pages} {P09011}
  (\bibinfo {year} {2012})}\BibitemShut {NoStop}%
\bibitem [{\citenamefont {P.Calabrese}\ \emph {et~al.}(2011)\citenamefont
  {P.Calabrese}, \citenamefont {Essler},\ and\ \citenamefont
  {Fagotti}}]{art:Calabrese(2011)}%
  \BibitemOpen
  \bibfield  {author} {\bibinfo {author} {\bibnamefont {P.Calabrese}}, \bibinfo
  {author} {\bibfnamefont {F.~H.~L.}\ \bibnamefont {Essler}}, \ and\ \bibinfo
  {author} {\bibfnamefont {M.}~\bibnamefont {Fagotti}},\ }\href@noop {}
  {\bibfield  {journal} {\bibinfo  {journal} {Phys. Rev. Lett.}\ }\textbf
  {\bibinfo {volume} {106}},\ \bibinfo {pages} {227203} (\bibinfo {year}
  {2011})}\BibitemShut {NoStop}%
\bibitem [{\citenamefont {P.Calabrese}(2012)}]{art:Calabrese(2012)}%
  \BibitemOpen
  \bibfield  {author} {\bibinfo {author} {\bibfnamefont {M.~F.}\ \bibnamefont
  {P.Calabrese}, \bibfnamefont {F.~H. L.~Essler}},\ }\href@noop {} {\bibfield
  {journal} {\bibinfo  {journal} {J. Stat. Mech.}\ }\textbf {\bibinfo {volume}
  {2012}},\ \bibinfo {pages} {P07016} (\bibinfo {year} {2012})}\BibitemShut
  {NoStop}%
\bibitem [{\citenamefont {P.Calabrese}\ \emph {et~al.}(2012)\citenamefont
  {P.Calabrese}, \citenamefont {Essler},\ and\ \citenamefont
  {Fagotti}}]{art:Calabrese(2012):2}%
  \BibitemOpen
  \bibfield  {author} {\bibinfo {author} {\bibnamefont {P.Calabrese}}, \bibinfo
  {author} {\bibfnamefont {F.~H.~L.}\ \bibnamefont {Essler}}, \ and\ \bibinfo
  {author} {\bibfnamefont {M.}~\bibnamefont {Fagotti}},\ }\href@noop {}
  {\bibfield  {journal} {\bibinfo  {journal} {J. Stat. Mech.}\ }\textbf
  {\bibinfo {volume} {2012}},\ \bibinfo {pages} {P07022} (\bibinfo {year}
  {2012})}\BibitemShut {NoStop}%
\bibitem [{\citenamefont {Essler}\ and\ \citenamefont
  {Fagotti}(2016)}]{art:Hessler(2016)}%
  \BibitemOpen
  \bibfield  {author} {\bibinfo {author} {\bibfnamefont {F.~H.~L.}\
  \bibnamefont {Essler}}\ and\ \bibinfo {author} {\bibfnamefont
  {M.}~\bibnamefont {Fagotti}},\ }\href@noop {} {\bibfield  {journal} {\bibinfo
   {journal} {J. Stat. Mech.}\ } (\bibinfo {year} {2016})}\BibitemShut
  {NoStop}%
\bibitem [{\citenamefont {Hess}\ \emph {et~al.}(2017)\citenamefont {Hess},
  \citenamefont {Becker}, \citenamefont {Kaplan}, \citenamefont {Kyprianidis},
  \citenamefont {Lee}, \citenamefont {Neyenhuis}, \citenamefont {Pagano},
  \citenamefont {Richerme}, \citenamefont {Senko}, \citenamefont {Smith} \emph
  {et~al.}}]{art:Hess(2017)}%
  \BibitemOpen
  \bibfield  {author} {\bibinfo {author} {\bibfnamefont {P.}~\bibnamefont
  {Hess}}, \bibinfo {author} {\bibfnamefont {P.}~\bibnamefont {Becker}},
  \bibinfo {author} {\bibfnamefont {H.}~\bibnamefont {Kaplan}}, \bibinfo
  {author} {\bibfnamefont {A.}~\bibnamefont {Kyprianidis}}, \bibinfo {author}
  {\bibfnamefont {A.}~\bibnamefont {Lee}}, \bibinfo {author} {\bibfnamefont
  {B.}~\bibnamefont {Neyenhuis}}, \bibinfo {author} {\bibfnamefont
  {G.}~\bibnamefont {Pagano}}, \bibinfo {author} {\bibfnamefont
  {P.}~\bibnamefont {Richerme}}, \bibinfo {author} {\bibfnamefont
  {C.}~\bibnamefont {Senko}}, \bibinfo {author} {\bibfnamefont
  {J.}~\bibnamefont {Smith}},  \emph {et~al.},\ }\href@noop {} {\bibfield
  {journal} {\bibinfo  {journal} {Philosophical Transactions of the Royal
  Society A: Mathematical, Physical and Engineering Sciences}\ }\textbf
  {\bibinfo {volume} {375}},\ \bibinfo {pages} {20170107} (\bibinfo {year}
  {2017})}\BibitemShut {NoStop}%
\bibitem [{\citenamefont {Rigol}\ \emph {et~al.}(2008)\citenamefont {Rigol},
  \citenamefont {Dunjko},\ and\ \citenamefont {Olshanii}}]{art:Rigol(2008)}%
  \BibitemOpen
  \bibfield  {author} {\bibinfo {author} {\bibfnamefont {M.}~\bibnamefont
  {Rigol}}, \bibinfo {author} {\bibfnamefont {V.}~\bibnamefont {Dunjko}}, \
  and\ \bibinfo {author} {\bibfnamefont {M.}~\bibnamefont {Olshanii}},\
  }\href@noop {} {\bibfield  {journal} {\bibinfo  {journal} {Nature}\ }\textbf
  {\bibinfo {volume} {452}},\ \bibinfo {pages} {854} (\bibinfo {year}
  {2008})}\BibitemShut {NoStop}%
\bibitem [{\citenamefont {Srednicki}(1994)}]{art:Srednicki(1994)}%
  \BibitemOpen
  \bibfield  {author} {\bibinfo {author} {\bibfnamefont {M.}~\bibnamefont
  {Srednicki}},\ }\href@noop {} {\bibfield  {journal} {\bibinfo  {journal}
  {Physical Review E}\ }\textbf {\bibinfo {volume} {50}},\ \bibinfo {pages}
  {888} (\bibinfo {year} {1994})}\BibitemShut {NoStop}%
\bibitem [{\citenamefont {Zhang}\ \emph {et~al.}(2017)\citenamefont {Zhang},
  \citenamefont {Hess}, \citenamefont {Kypriandis}, \citenamefont {Becker},
  \citenamefont {Lee}, \citenamefont {Smith}, \citenamefont {Pagano},
  \citenamefont {Ptorniche}, \citenamefont {Potter}, \citenamefont
  {Vishwanath}, \citenamefont {Yao},\ and\ \citenamefont
  {Monroe}}]{art:Zhang(2017)}%
  \BibitemOpen
  \bibfield  {author} {\bibinfo {author} {\bibfnamefont {J.}~\bibnamefont
  {Zhang}}, \bibinfo {author} {\bibfnamefont {P.~W.}\ \bibnamefont {Hess}},
  \bibinfo {author} {\bibfnamefont {A.}~\bibnamefont {Kypriandis}}, \bibinfo
  {author} {\bibfnamefont {P.}~\bibnamefont {Becker}}, \bibinfo {author}
  {\bibfnamefont {A.}~\bibnamefont {Lee}}, \bibinfo {author} {\bibfnamefont
  {J.}~\bibnamefont {Smith}}, \bibinfo {author} {\bibfnamefont
  {G.}~\bibnamefont {Pagano}}, \bibinfo {author} {\bibfnamefont {I.-D.}\
  \bibnamefont {Ptorniche}}, \bibinfo {author} {\bibfnamefont {A.~C.}\
  \bibnamefont {Potter}}, \bibinfo {author} {\bibfnamefont {A.}~\bibnamefont
  {Vishwanath}}, \bibinfo {author} {\bibfnamefont {N.~Y.}\ \bibnamefont {Yao}},
  \ and\ \bibinfo {author} {\bibfnamefont {C.}~\bibnamefont {Monroe}},\
  }\href@noop {} {\bibfield  {journal} {\bibinfo  {journal} {Nature}\ }\textbf
  {\bibinfo {volume} {543}},\ \bibinfo {pages} {217} (\bibinfo {year}
  {2017})}\BibitemShut {NoStop}%
\bibitem [{\citenamefont {Gambassi}\ and\ \citenamefont
  {Calabrese}(2011)}]{art:Gambassi_C(2011)}%
  \BibitemOpen
  \bibfield  {author} {\bibinfo {author} {\bibfnamefont {A.}~\bibnamefont
  {Gambassi}}\ and\ \bibinfo {author} {\bibfnamefont {P.}~\bibnamefont
  {Calabrese}},\ }\href@noop {} {\bibfield  {journal} {\bibinfo  {journal} {EPL
  (Europhysics Letters)}\ }\textbf {\bibinfo {volume} {95}},\ \bibinfo {pages}
  {66007} (\bibinfo {year} {2011})}\BibitemShut {NoStop}%
\bibitem [{\citenamefont {Sciolla}\ and\ \citenamefont
  {Biroli}(2010)}]{art:Sciolla(2010)}%
  \BibitemOpen
  \bibfield  {author} {\bibinfo {author} {\bibfnamefont {B.}~\bibnamefont
  {Sciolla}}\ and\ \bibinfo {author} {\bibfnamefont {G.}~\bibnamefont
  {Biroli}},\ }\href@noop {} {\bibfield  {journal} {\bibinfo  {journal}
  {Physical review letters}\ }\textbf {\bibinfo {volume} {105}},\ \bibinfo
  {pages} {220401} (\bibinfo {year} {2010})}\BibitemShut {NoStop}%
\bibitem [{\citenamefont {Heyl}\ \emph {et~al.}(2013)\citenamefont {Heyl},
  \citenamefont {Polkovnikov},\ and\ \citenamefont {Kehrein}}]{art:Heyl(2013)}%
  \BibitemOpen
  \bibfield  {author} {\bibinfo {author} {\bibfnamefont {M.}~\bibnamefont
  {Heyl}}, \bibinfo {author} {\bibfnamefont {A.}~\bibnamefont {Polkovnikov}}, \
  and\ \bibinfo {author} {\bibfnamefont {S.}~\bibnamefont {Kehrein}},\
  }\href@noop {} {\bibfield  {journal} {\bibinfo  {journal} {Phys. Rev. Lett.}\
  }\textbf {\bibinfo {volume} {110}},\ \bibinfo {pages} {135704} (\bibinfo
  {year} {2013})}\BibitemShut {NoStop}%
\bibitem [{\citenamefont {Heyl}(2014)}]{art:Heyl(2014)}%
  \BibitemOpen
  \bibfield  {author} {\bibinfo {author} {\bibfnamefont {M.}~\bibnamefont
  {Heyl}},\ }\href@noop {} {\bibfield  {journal} {\bibinfo  {journal} {Physical
  review letters}\ }\textbf {\bibinfo {volume} {113}},\ \bibinfo {pages}
  {205701} (\bibinfo {year} {2014})}\BibitemShut {NoStop}%
\bibitem [{\citenamefont {Jurcevic}\ \emph {et~al.}(2017)\citenamefont
  {Jurcevic}, \citenamefont {Shen}, \citenamefont {Hauke}, \citenamefont
  {Maier}, \citenamefont {Brydges}, \citenamefont {Hempel}, \citenamefont
  {Lanyon}, \citenamefont {Heyl}, \citenamefont {Blatt},\ and\ \citenamefont
  {Roos}}]{art:Jurcevic(2017)}%
  \BibitemOpen
  \bibfield  {author} {\bibinfo {author} {\bibfnamefont {P.}~\bibnamefont
  {Jurcevic}}, \bibinfo {author} {\bibfnamefont {H.}~\bibnamefont {Shen}},
  \bibinfo {author} {\bibfnamefont {P.}~\bibnamefont {Hauke}}, \bibinfo
  {author} {\bibfnamefont {C.}~\bibnamefont {Maier}}, \bibinfo {author}
  {\bibfnamefont {T.}~\bibnamefont {Brydges}}, \bibinfo {author} {\bibfnamefont
  {C.}~\bibnamefont {Hempel}}, \bibinfo {author} {\bibfnamefont
  {B.}~\bibnamefont {Lanyon}}, \bibinfo {author} {\bibfnamefont
  {M.}~\bibnamefont {Heyl}}, \bibinfo {author} {\bibfnamefont {R.}~\bibnamefont
  {Blatt}}, \ and\ \bibinfo {author} {\bibfnamefont {C.}~\bibnamefont {Roos}},\
  }\href@noop {} {\bibfield  {journal} {\bibinfo  {journal} {Physical review
  letters}\ }\textbf {\bibinfo {volume} {119}},\ \bibinfo {pages} {080501}
  (\bibinfo {year} {2017})}\BibitemShut {NoStop}%
\bibitem [{\citenamefont {{\v{Z}}unkovi{\v{c}}}\ \emph
  {et~al.}(2018)\citenamefont {{\v{Z}}unkovi{\v{c}}}, \citenamefont {Heyl},
  \citenamefont {Knap},\ and\ \citenamefont {Silva}}]{art:Knap(2018)}%
  \BibitemOpen
  \bibfield  {author} {\bibinfo {author} {\bibfnamefont {B.}~\bibnamefont
  {{\v{Z}}unkovi{\v{c}}}}, \bibinfo {author} {\bibfnamefont {M.}~\bibnamefont
  {Heyl}}, \bibinfo {author} {\bibfnamefont {M.}~\bibnamefont {Knap}}, \ and\
  \bibinfo {author} {\bibfnamefont {A.}~\bibnamefont {Silva}},\ }\href@noop {}
  {\bibfield  {journal} {\bibinfo  {journal} {Physical review letters}\
  }\textbf {\bibinfo {volume} {120}},\ \bibinfo {pages} {130601} (\bibinfo
  {year} {2018})}\BibitemShut {NoStop}%
\bibitem [{\citenamefont {Lacki}\ and\ \citenamefont
  {Heyl}(2018)}]{art:Lacki(2018)}%
  \BibitemOpen
  \bibfield  {author} {\bibinfo {author} {\bibfnamefont {M.}~\bibnamefont
  {Lacki}}\ and\ \bibinfo {author} {\bibfnamefont {M.}~\bibnamefont {Heyl}},\
  }\href@noop {} {\bibfield  {journal} {\bibinfo  {journal} {arXiv preprint
  arXiv:1812.02209}\ } (\bibinfo {year} {2018})}\BibitemShut {NoStop}%
\bibitem [{\citenamefont {Usui}\ \emph {et~al.}(2017)\citenamefont {Usui},
  \citenamefont {Busch}, \citenamefont {Silva}, \citenamefont {Goold} \emph
  {et~al.}}]{art:Fogarty(2017)}%
  \BibitemOpen
  \bibfield  {author} {\bibinfo {author} {\bibfnamefont {A.}~\bibnamefont
  {Usui}}, \bibinfo {author} {\bibfnamefont {T.}~\bibnamefont {Busch}},
  \bibinfo {author} {\bibfnamefont {A.}~\bibnamefont {Silva}}, \bibinfo
  {author} {\bibfnamefont {J.}~\bibnamefont {Goold}},  \emph {et~al.},\
  }\href@noop {} {\bibfield  {journal} {\bibinfo  {journal} {New Journal of
  Physics}\ }\textbf {\bibinfo {volume} {19}},\ \bibinfo {pages} {113018}
  (\bibinfo {year} {2017})}\BibitemShut {NoStop}%
\bibitem [{\citenamefont {Heyl}(2019)}]{art:Heyl(2018)}%
  \BibitemOpen
  \bibfield  {author} {\bibinfo {author} {\bibfnamefont {M.}~\bibnamefont
  {Heyl}},\ }\href@noop {} {\bibfield  {journal} {\bibinfo  {journal} {EPL
  (Europhysics Letters)}\ }\textbf {\bibinfo {volume} {125}},\ \bibinfo {pages}
  {26001} (\bibinfo {year} {2019})}\BibitemShut {NoStop}%
\bibitem [{\citenamefont {Halimeh}\ and\ \citenamefont
  {Zauner-Stauber}(2017{\natexlab{a}})}]{art:Halimeh(2017)}%
  \BibitemOpen
  \bibfield  {author} {\bibinfo {author} {\bibfnamefont {J.~C.}\ \bibnamefont
  {Halimeh}}\ and\ \bibinfo {author} {\bibfnamefont {V.}~\bibnamefont
  {Zauner-Stauber}},\ }\href@noop {} {\bibfield  {journal} {\bibinfo  {journal}
  {Physical Review B}\ }\textbf {\bibinfo {volume} {96}},\ \bibinfo {pages}
  {134427} (\bibinfo {year} {2017}{\natexlab{a}})}\BibitemShut {NoStop}%
\bibitem [{\citenamefont {Halimeh}\ and\ \citenamefont
  {Zauner-Stauber}(2017{\natexlab{b}})}]{art:Halimeh(2017)2}%
  \BibitemOpen
  \bibfield  {author} {\bibinfo {author} {\bibfnamefont {J.~C.}\ \bibnamefont
  {Halimeh}}\ and\ \bibinfo {author} {\bibfnamefont {V.}~\bibnamefont
  {Zauner-Stauber}},\ }\href@noop {} {\bibfield  {journal} {\bibinfo  {journal}
  {Physical Review B}\ }\textbf {\bibinfo {volume} {96}},\ \bibinfo {pages}
  {134427} (\bibinfo {year} {2017}{\natexlab{b}})}\BibitemShut {NoStop}%
\bibitem [{\citenamefont {Lang}\ \emph {et~al.}(2018)\citenamefont {Lang},
  \citenamefont {Frank},\ and\ \citenamefont {Halimeh}}]{art:Halimeh(2018)}%
  \BibitemOpen
  \bibfield  {author} {\bibinfo {author} {\bibfnamefont {J.}~\bibnamefont
  {Lang}}, \bibinfo {author} {\bibfnamefont {B.}~\bibnamefont {Frank}}, \ and\
  \bibinfo {author} {\bibfnamefont {J.~C.}\ \bibnamefont {Halimeh}},\
  }\href@noop {} {\bibfield  {journal} {\bibinfo  {journal} {Physical review
  letters}\ }\textbf {\bibinfo {volume} {121}},\ \bibinfo {pages} {130603}
  (\bibinfo {year} {2018})}\BibitemShut {NoStop}%
\bibitem [{\citenamefont {Rossini}\ \emph {et~al.}(2014)\citenamefont
  {Rossini}, \citenamefont {Fazio}, \citenamefont {Giovannetti},\ and\
  \citenamefont {Silva}}]{art:Rossini(2014)}%
  \BibitemOpen
  \bibfield  {author} {\bibinfo {author} {\bibfnamefont {D.}~\bibnamefont
  {Rossini}}, \bibinfo {author} {\bibfnamefont {R.}~\bibnamefont {Fazio}},
  \bibinfo {author} {\bibfnamefont {V.}~\bibnamefont {Giovannetti}}, \ and\
  \bibinfo {author} {\bibfnamefont {A.}~\bibnamefont {Silva}},\ }\href@noop {}
  {\bibfield  {journal} {\bibinfo  {journal} {Europhys. Lett.}\ }\textbf
  {\bibinfo {volume} {107}},\ \bibinfo {pages} {30002} (\bibinfo {year}
  {2014})}\BibitemShut {NoStop}%
\bibitem [{\citenamefont {Barouch}\ \emph {et~al.}(1970)\citenamefont
  {Barouch}, \citenamefont {McCoy},\ and\ \citenamefont
  {Dresden}}]{art:Barouch(1970)}%
  \BibitemOpen
  \bibfield  {author} {\bibinfo {author} {\bibfnamefont {E.}~\bibnamefont
  {Barouch}}, \bibinfo {author} {\bibfnamefont {B.~M.}\ \bibnamefont {McCoy}},
  \ and\ \bibinfo {author} {\bibfnamefont {M.}~\bibnamefont {Dresden}},\
  }\href@noop {} {\bibfield  {journal} {\bibinfo  {journal} {Phys. Rev. A}\
  }\textbf {\bibinfo {volume} {2}},\ \bibinfo {pages} {1075} (\bibinfo {year}
  {1970})}\BibitemShut {NoStop}%
\bibitem [{\citenamefont {Barouch}\ \emph {et~al.}(1971)\citenamefont
  {Barouch}, \citenamefont {McCoy},\ and\ \citenamefont
  {Dresden}}]{art:Barouch(1971)}%
  \BibitemOpen
  \bibfield  {author} {\bibinfo {author} {\bibfnamefont {E.}~\bibnamefont
  {Barouch}}, \bibinfo {author} {\bibfnamefont {B.~M.}\ \bibnamefont {McCoy}},
  \ and\ \bibinfo {author} {\bibfnamefont {M.}~\bibnamefont {Dresden}},\
  }\href@noop {} {\bibfield  {journal} {\bibinfo  {journal} {Phys. Rev. A}\
  }\textbf {\bibinfo {volume} {3}},\ \bibinfo {pages} {786} (\bibinfo {year}
  {1971})}\BibitemShut {NoStop}%
\bibitem [{\citenamefont {Sachdev}(2000)}]{book:Sachdev(2000)}%
  \BibitemOpen
  \bibfield  {author} {\bibinfo {author} {\bibfnamefont {S.}~\bibnamefont
  {Sachdev}},\ }\href@noop {} {\emph {\bibinfo {title} {Quantum phase
  transition}}}\ (\bibinfo  {publisher} {Cambridge University Press},\ \bibinfo
  {year} {2000})\BibitemShut {NoStop}%
\bibitem [{\citenamefont {Allahverdyan A.~E.}\ and\ \citenamefont
  {M.}(2013)}]{art:Allahverdyan(2004)}%
  \BibitemOpen
  \bibfield  {author} {\bibinfo {author} {\bibfnamefont {B.~R.}\ \bibnamefont
  {Allahverdyan A.~E.}}\ and\ \bibinfo {author} {\bibfnamefont {N.~T.}\
  \bibnamefont {M.}},\ }\href@noop {} {\bibfield  {journal} {\bibinfo
  {journal} {Europhys. Lett.}\ }\textbf {\bibinfo {volume} {67}},\ \bibinfo
  {pages} {565} (\bibinfo {year} {2013})}\BibitemShut {NoStop}%
\bibitem [{\citenamefont {Russomanno}\ \emph {et~al.}(2013)\citenamefont
  {Russomanno}, \citenamefont {Silva},\ and\ \citenamefont
  {Santoro}}]{art:Russomanno(2013)}%
  \BibitemOpen
  \bibfield  {author} {\bibinfo {author} {\bibfnamefont {A.}~\bibnamefont
  {Russomanno}}, \bibinfo {author} {\bibfnamefont {A.}~\bibnamefont {Silva}}, \
  and\ \bibinfo {author} {\bibfnamefont {G.}~\bibnamefont {Santoro}},\
  }\href@noop {} {\bibfield  {journal} {\bibinfo  {journal} {J. Stat. Mech.}\
  }\textbf {\bibinfo {volume} {2013}},\ \bibinfo {pages} {P09012} (\bibinfo
  {year} {2013})}\BibitemShut {NoStop}%
\end{thebibliography}%
\end{document}